\documentclass[nofootinbib, aps, prc]{revtex4-1}

\usepackage{epsfig}
\usepackage[utf8]{inputenc}
\usepackage{amsmath}
\usepackage{amssymb}
\usepackage{amsthm}
\usepackage{amsfonts}
\usepackage{xspace}
\usepackage{graphicx}
\usepackage{graphics}
\usepackage{slashed}
\usepackage{bbold}
\usepackage{bbm}
\usepackage{physics}
\usepackage{empheq}
\usepackage{mathrsfs}
\usepackage{xcolor}

\usepackage{comment}

\usepackage{cancel}

\usepackage{hyperref}

\newcommand{\Nc}{\ensuremath{N_c}\xspace}
\newcommand{\LONc}{\ensuremath{\text{LO-in-\Nc}}\xspace}
\newcommand{\NLONc}{\ensuremath{\text{NLO-in-\Nc}}\xspace}

\newcommand{\NN}{\ensuremath{N\!N}\xspace}

\newcommand{\neutrinoless}{\ensuremath{0 \nu \beta \beta}\xspace}

\newcommand{\oneS}{{{}^{1}\!S_0}}

\newcommand{\calO}{\ensuremath{\mathcal{O}}}

\newcommand{\calC}{\ensuremath{\mathcal{C}}}

\newcommand{\Lagr}{\mathcal{L}}

\newcommand{\1}{\mathbbm{1}}

\renewcommand\vector{\mathbf}

\newcommand{\pplus}{\vector{p}_+}
\newcommand{\pminus}{\vector{p}_-}

\newcommand{\gnu}{\ensuremath{g_\nu^{\NN}}\xspace}

\newcommand{\nopi}{\ensuremath{\pi\hskip-0.40em /}}
\newcommand{\eftnopi}{EFT$_{\nopi}$\xspace}

\newcommand{\chiPT}{$\chi$PT\xspace}

\begin{document}

\title{Large-\Nc analysis of two-nucleon neutrinoless double beta decay and charge-independence-breaking contact terms}

\author{Thomas R.~Richardson}
\email{richa399@email.sc.edu}
\author{Matthias R.~Schindler}
\email{mschindl@mailbox.sc.edu}
\affiliation{Department of Physics and Astronomy,
University of South Carolina, 
Columbia, SC 29208}
\author{Saori Pastore}
\email{saori@wustl.edu}
\affiliation{Department of Physics and McDonnell Center for the Space Sciences at Washington University in St. Louis, MO, 63130, USA}
\author{Roxanne P. Springer}
\email{rps@phy.duke.edu}
\affiliation{Department  of  Physics,  Duke  University,  Durham,  NC  27708,  USA}

\date{\today}

\begin{abstract}
The interpretation of experiments that search for neutrinoless double beta decay relies on input from nuclear theory.
Cirigliano {\it et al.} recently showed that, for the light Majorana exchange formalism, effective field theory calculations require a $nn\to pp e^- e^-$ contact term at leading order. They estimated the size of this contribution  by relating it to measured charge-independence-breaking (CIB) nucleon-nucleon interactions and making an assumption about the relative sizes of CIB operators.  We show that the assumptions underlying this approximation are justified in the limit of the number of colors \Nc being large. We also obtain a large-\Nc hierarchy among CIB nucleon-nucleon interactions that is in agreement with phenomenological results.
\end{abstract}

\maketitle


\section{Introduction}
    \label{introduction}
    
Significant experimental efforts are underway to detect neutrinoless double beta ($\neutrinoless$) decay~\cite{PhysRevLett.117.109903,Albert:2014awa,Adams:2020cye,Zsigmond:2020bfx,Cattadori:2015xua,PhysRevLett.115.102502,Caden:2017htb,Blot:2016cei,Giuliani:2017eje,Tetsuno:2020ngo,PARK20162630,Ebert:2015pgx,Dokania:2015apa,Fukuda_2016}, a process in which two neutrons are converted into two protons with the emission of two electrons but without the accompanying emission of neutrinos. Neutrinoless double beta decay is a highly sensitive probe of 
lepton number violation (LNV) and, if detected, would be a clear demonstration that neutrinos are Majorana particles \cite{schechter_neutrinoless_1982, Zeldovich:1981da:english, *Zeldovich:1981da:russian}.
If this process is observed, it would also shed light on the neutrino mass hierarchy \cite{vergados_neutrinoless_2016, pas_neutrinoless_2015} and on the matter-antimatter asymmetry in the universe \cite{davidson_leptogenesis_2008}.

The inverse of the \neutrinoless half-life can be expressed as (see Refs.~\cite{engel_status_2017, drischler_towards_2020, avignone_double_2008} for reviews)
    \begin{equation}
        \left[ T_{1/2}^{0\nu} \right]^{-1} = G_{0\nu} \abs{M_{0 \nu}}^2 m_{\beta \beta}^2 \, ,
    \end{equation}
where $m_{\beta \beta}$ is the effective Majorana neutrino mass, $G_{0\nu}$ is a phase space factor, and $M_{0 \nu}$ is the corresponding nuclear matrix element (NME). 
This sensitivity to the NME requires a deep  understanding of the nuclear physics involved. 
One important component in the calculation of the NMEs are multi-nucleon operators that encode the underlying LNV mechanisms.
While there are many models and methods that offer insight in this direction, effective field theory (EFT) offers a systematic, model-independent way to study LNV and the corresponding one- and two-nucleon operators that are required as input to many-body calculations. Each independent term in an effective Lagrangian comes with a low-energy coefficient (LEC), into which all unresolved short-distance details are subsumed. These LECs need to be determined from a fit to data
or a nonperturbative quantum chromodynamics (QCD) calculation like those performed in lattice QCD (see, e.g., Refs.~\cite{tiburzi_double-_2017,Cirigliano:2020yhp,detmold_neutrinoless_2020,Cirigliano:2020yhp,Davoudi:2020xdv,Davoudi:2020gxs} for work related to double beta decays and Ref.~\cite{Davoudi:2020ngi} for a general review).

An initial step towards the application of EFT to \neutrinoless was taken in Ref.~\cite{prezeau_neutrinoless_2003} in the context of chiral effective field theory. ChEFT refers to the generalization of chiral perturbation theory (\chiPT) \cite{Weinberg:1978kz,gasser_chiral_1984,Gasser:1984gg,Gasser:1987rb,Jenkins:1990jv,Jenkins:1991ne} (see e.g., Refs.~\cite{Scherer:2002tk,Bijnens:2006zp,Bernard:2006gx,Bernard:2007zu,Birse:2007zz,Scherer:2012xha} for reviews)---the EFT of pions and single nucleons based on the approximate chiral symmetry of QCD---to two and more nucleons. 
Recently, the EFT approach has received renewed attention focusing on the light-Majorana neutrino exchange mechanism \cite{cirigliano_neutrinoless_2017,cirigliano_neutrinoless_2018,Cirigliano:2017ymo,Cirigliano:2018yza}, where it has been shown that a contact term with the undetermined LEC $\gnu$ is required at leading order (LO) \cite{cirigliano_new_2018,cirigliano_renormalized_2019}.
This term was absent in previous analyses. Additionally, Refs.~\cite{cirigliano_neutrinoless_2018,cirigliano_new_2018,cirigliano_renormalized_2019} observed that isospin symmetry dictates that $\gnu$ is  related to the LEC $\calC_1$ of an operator parameterizing charge independence breaking (CIB) in the two-nucleon system. 
The value of the LEC $\calC_1$ is currently not determined by data. Only the linear combination $\calC_1+\calC_2$, where $\calC_2$ is the LEC of a second, independent CIB operator, has been extracted from experiment. To estimate the numerical impact of the contact term in nuclear  matrix elements, Refs.~\cite{cirigliano_new_2018,cirigliano_renormalized_2019} assumed that $\calC_1\approx \calC_2$ so that the value of \gnu can be approximated by  $\gnu\approx \frac{1}{2}(\calC_1+\calC_2)$. Exploring this assumption from the large-\Nc perspective is the main focus of this paper.

While some lattice QCD calculations of double-$\beta$ decay matrix elements in the two-nucleon system \cite{tiburzi_double-_2017,Cirigliano:2020yhp} and \neutrinoless calculations in the meson sector \cite{detmold_neutrinoless_2020,Cirigliano:2020yhp} exist, a calculation of \gnu is currently not available.
In the absence of lattice QCD calculations and sufficient data to determine \gnu, or equivalently the CIB LECs, the possibility of additional theoretical constraints is critical.
Recently, Refs.~\cite{Cirigliano:2020dmx, Cirigliano:2021qko} estimated the values of \gnu and $\calC_1+\calC_2$ using a method analogous to the Cottingham formula \cite{Cottingham:1963zz,Harari:1966mu}. Their results support the assumption of Refs.~\cite{cirigliano_new_2018,cirigliano_renormalized_2019}.
Here, a complementary approach based on the large-\Nc limit of QCD is explored. Constraints are obtained through the spin-flavor symmetry that arises in the large-\Nc limit of QCD \cite{Dashen:1993as,Jenkins:1993af,dashen_1/nc_1994,dashen_spin_1995}. 
This method has been used to constrain nucleon-nucleon (\NN) interactions \cite{kaplan_spin_1996, kaplan_nucleon-nucleon_1997, banerjee_nucleon_2002,Riska:2002vn, schindler_large-$n_c$_2018}, including parity-violating couplings \cite{schindler_large-$n_c$_2016, phillips_parity-violating_2014}, time-reversal-invariance-violating couplings \cite{vanasse_time-reversal-invariance_2019, samart_time-reversal_2016}, as well as magnetic and axial couplings in the context of pionless EFT (\eftnopi) \cite{richardson_large-_2020}. Similar work has been done in both the meson \cite{leutwyler_bounds_1996, kaiser_large_2000} and single baryon sectors of chiral perturbation theory (\chiPT) \cite{jenkins_chiral_1996, calle_cordon_baryon_2013, flores-mendieta_baryon_2014, flores-mendieta_structure_2000, bedaque_baryon_1996}.

There are potential pitfalls to applying  the large-\Nc expansion to nuclei. For example, Skyrme models suggests that the binding energy per nucleon in nuclear matter predicted in the large-\Nc limit is of order the nucleon mass $m_N\sim\Nc$ \cite{Kutschera:1984zm,Klebanov:1985qi}, while the observed binding energies are much smaller (of the order of a few MeVs.) For a more detailed discussion see, e.g., Refs.~\cite{McLerran:2009nr,Hidaka:2010ph} and references therein. 
Here, we use arguments based on the spin-flavor symmetry to compare the relative sizes of different terms in the \NN Lagrangian. This approach gives results for the isospin-invariant \NN interactions that are consistent with \NN scattering data \cite{kaplan_spin_1996,kaplan_nucleon-nucleon_1997,schindler_large-$n_c$_2018}.
Moreover, the Wigner symmetry that was shown to emerge in the large-\Nc limit yields   agreement with some parity conserving experimental results for larger nuclei (see Ref.~\cite{kaplan_spin_1996} and references therein).

This paper is structured as follows.
Section \ref{Sec:Background} contains a discussion of the results from Refs.~\cite{cirigliano_new_2018, cirigliano_renormalized_2019} relevant for this work. A large-\Nc analysis of one- and two-nucleon matrix elements is given in Sec.~\ref{LNEFT}, and the spurion construction in ChEFT is discussed in Sec.~\ref{Sec:Spurion}. Complete but minimal sets of spurion operators for both the electromagnetic and weak interactions are derived along with the large-\Nc scalings of the corresponding LECs in Sec.~\ref{Sec:LargeN_scaling}. A large-\Nc hierarchy of CIB interactions in comparison to phenomenological descriptions is discussed in  Sec.~\ref{Sec:hierarchy}. The large-\Nc analyzed CIB Lagrangian is mapped onto to the CIB Lagrangian of Ref.~\cite{cirigliano_renormalized_2019} in Sec.~\ref{Sec:Justification} and the consistency of the LNV and CIB LECs demonstrated. Finally, Sec.~\ref{Sec:conclusion} summarizes the results.  The appendices contain a detailed discussion of an alternate large-\Nc scaling of the quark and nucleon charges, as well as a summary of relevant Fierz identities.


\section{Background}
    \label{Sec:Background}

In this section we introduce and discuss the relevant LNV and CIB  Lagrangians.
At leading order (LO) in the EFT power counting, there is a contribution to the two-nucleon LNV transition operator from  tree-level neutrino exchange between the nucleons. 
 At the same order, there exist contributions from dressing the tree-level diagram by iterations of the LO \NN interactions, which include  contact terms and one-pion exchange diagrams \cite{cirigliano_renormalized_2019}. A careful analysis~\cite{cirigliano_new_2018,cirigliano_renormalized_2019} of the resulting amplitude using renormalization arguments shows that an LNV amplitude that consists of only the above contributions diverges logarithmically.
 Therefore, a leading-order contact operator must be included to obtain the correct amplitude at this order. 
The contact term in the LO Lagrangian is \cite{cirigliano_new_2018,cirigliano_renormalized_2019}
    \begin{equation}
            \label{CiriglianoLNV}
        \Lagr^{\NN}_{ \abs{\Delta L = 2} } = \left( 2 \sqrt{2} G_F V_{ud} \right)^2 m_{\beta \beta} \bar e_L C \bar e^T_L \frac{g^{NN}_\nu}{4} \left[ \left( \bar N u \tilde Q_L^w u^\dagger N \right)^2 - \frac{1}{6} \Tr( \tilde Q_L^w \tilde Q_L^w ) \left( \bar N \tau^a N \right)^2 \right] +\text{H.c.},
    \end{equation}
where $N$ represents the doublet of nucleon fields, $e_L$ is the left-handed electron, the charge conjugation matrix is $C = i \gamma^2 \gamma^0$, $G_F$ is the Fermi constant, $V_{ud}$ is an element of the Cabibbo-Kobayashi-Maskawa matrix, and
\begin{equation}
    \tilde Q_L^w = \tau^+ = \frac{1}{2} \left( \tau^1 + i \tau^2 \right)\ .
    \label{eq:weakspurion}
\end{equation}
The matrix $u$ is
    \begin{equation}
        u = \text{exp} \left(\frac{i}{2F} \phi_a \tau^a \right),
        \label{eq:udef}
    \end{equation}
where the $\phi_a$ $(a=1,2,3)$  are the pion fields in Cartesian coordinates, the $\tau^a$ are Pauli matrices in isospin space, and $F$ is the pion decay constant in the chiral limit.
The renormalization group (RG) requirement to include a contact term at LO means that an additional   unknown LEC, \gnu, must be determined in order to analyze and interpret current and future measurements of \neutrinoless decay.

It has been shown in Refs.~\cite{cirigliano_new_2018,cirigliano_renormalized_2019} that chiral symmetry relates the LEC \gnu to an electromagnetic CIB  isotensor LEC $\calC_1$.
The CIB isotensor Lagrangian in ChEFT has received a significant amount of study \cite{epelbaum_charge_1999, walzl_charge-dependent_2001, van_kolck_soft_1993}. In Ref.~\cite{cirigliano_renormalized_2019} it is written as 
    \begin{eqnarray}
        \Lagr_{CIB}^{\NN} & = & \frac{e^2}{4} \left\{ \calC_1 \left[ \left( \bar N u^\dagger \tilde Q_R u N \right)^2 + \left( \bar N u \tilde Q_L u^\dagger N \right)^2 - \frac{1}{6} \Tr( \tilde Q_L^2 + \tilde Q_R^2 ) \left(\bar N \tau^a N \right)^2 \right] \right. \nonumber \\
        & & \left. +  \calC_2 \left[  2 \left( \bar N u^\dagger \tilde Q_R u N \right) \left( \bar N u \tilde Q_L u^\dagger N \right) - \frac{1}{3} \Tr( U \tilde Q_L U^\dagger \tilde Q_R ) \left( \bar N \tau^a N \right)^2  \right] \right\}, \label{CiriglianoCIB:1}  
    \end{eqnarray}
where $U=u^2$ and here
    \begin{equation}
    \tilde Q_L = \tilde Q_R = \frac{1}{2} \tau^3.
    \label{eq:EMspurion}
\end{equation}

Additionally, many high-precision \NN potentials, such as the Argonne $v_{18}$ \cite{wiringa_accurate_1995} and the CD-Bonn \cite{machleidt_precision_2001}, as well as several interactions derived from ChEFT \cite{Piarulli:2016vel,Epelbaum:2014efa,Reinert:2017usi,Machleidt:2011zz} include short-range CIB and charge-symmetry-breaking (CSB) operators to reproduce the observed scattering data.
In principle, determination of $\calC_1$ from data also fixes the value of the \neutrinoless LEC \gnu. However, at present only the linear combination $\calC_1+\calC_2$ is constrained by available data. The combination $\calC_1 - \calC_2$ is sensitive to two-nucleon-multi-pion interactions and is currently inaccessible.
Reference~\cite{cirigliano_renormalized_2019} obtains an estimate of \gnu by assuming that the two LECs $\calC_1$ and $\calC_2$ are of the same size and sign, which implies  $g_\nu^{NN} \approx \frac{1}{2}(\calC_1+\calC_2)$. In the next sections we examine this assumption using large-\Nc scaling arguments.  
In particular, we show that the terms proportional to $\calC_1 - \calC_2$ are suppressed in the large-\Nc limit compared to those proportional to $\calC_1 + \calC_2$, thereby adding support to the assumptions that $\calC_1$ and $\calC_2$ are of the same size and sign and that \gnu can be approximated by the sum of the CIB LECs divided by two.


\section{Large-\Nc scaling}
    \label{LNEFT}

In this section we outline the basic elements needed to perform large-\Nc analyses.
 The large-\Nc scaling of the single-nucleon matrix elements of an $n$-body operator $\calO^{(n)}_{IS}$ with spin $S$ and isospin $I$ is provided by \cite{kaplan_spin_1996, kaplan_nucleon-nucleon_1997}
    \begin{equation}
            \label{Eq:OneBodyScaling}
        \bra{N} \frac{ \calO^{(n)}_{IS} }{ \Nc^n } \ket{N} \lesssim \Nc^{- \abs{I - S}},
    \end{equation}
where $n$ denotes the number of quarks involved in the operator. In the large-\Nc limit, the Hamiltonian takes a Hartree form
\cite{kaplan_nucleon-nucleon_1997, witten_baryons_1979},
    \begin{equation}
        H = \Nc \sum_{n} \sum_{s,t} v_{stn} \left( \frac{ {S}^i }{N_c} \right)^s \left( \frac{ {I}^a }{N_c} \right)^t \left( \frac{ {G}^{ia} }{N_c} \right)^{n-s-t},
        \label{Hartree}
    \end{equation}
where the one-body operators are
    \begin{equation}
            \label{OneBodyLargeN}
        {S}^i = q^\dagger \frac{\sigma^i}{2}q, \ \ {I}^a = q^\dagger \frac{\tau^a}{2} q, \ \ {G}^{ia} = q^\dagger \frac{\sigma^i \tau^a}{4} q.
    \end{equation}
  The nucleon ground state is totally antisymmetric in the color degrees of freedom, and $q$ is a colorless, bosonic quark field.  
The coefficients $v_{stn}$ are functions of momentum and at most scale as $O(\Nc^0)$ \cite{kaplan_nucleon-nucleon_1997}.
In addition to single-nucleon
matrix elements (also see Ref.~\cite{jenkins_large-n_c_1998} and references therein), these results were used in the study of two-nucleon interactions via matrix elements of the form \cite{kaplan_nucleon-nucleon_1997, kaplan_spin_1996}
    \begin{equation}
       V(\pminus,\pplus) =  \bra{N_\alpha(\vector{p}^\prime_1) N_\beta(\vector{p}^\prime_2)} H \ket{N_\gamma(\vector{p}_1) N_\delta(\vector{p}_2)},
    \end{equation}
where the Greek subscripts indicate combined spin and isospin quantum numbers and 
    \begin{equation}
      \vector{p}_\pm \equiv \vector{p}^\prime \pm \vector{p},    
    \end{equation}
where $\vector{p}^\prime = \vector{p}^\prime_1 - \vector{p}^\prime_2$ and $\vector{p} = \vector{p}_1 - \vector{p}_2$. The two-nucleon matrix elements factorize in the large-\Nc limit \cite{kaplan_spin_1996}, 
    \begin{equation}
        \bra{N_\gamma N_\delta} \mathcal{O}_1 \mathcal{O}_2 \ket{N_\alpha N_\beta} \xrightarrow[]{N_c \to \infty} \bra{N_\gamma} \mathcal O_1 \ket{N_\alpha} \bra{N_\delta} \mathcal O_2 \ket{N_\beta} + \text{crossed},
    \end{equation}
and the large-\Nc scaling of the two-nucleon matrix elements is determined by the large-\Nc dependence of the operators  $\calO_{1}$ and $\calO_{2}$  $\in \{S^i, I^a,G^{ia},\1\}$, 
    \begin{equation}
    \begin{split}
        &\bra{N'} S^i \ket{N} \sim \bra{N} I^a \ket{N} \lesssim 1 \, , \\
         &\bra{N'} G^{ia} \ket{N} \sim \bra{N} \1 \ket{N} \lesssim \Nc\ .
        \end{split}
        \label{MEScaling}
    \end{equation}

In addition, there can be a hidden large-\Nc suppression in the momentum dependence of the functions $v_{stn}$ \cite{kaplan_nucleon-nucleon_1997}. In t-channel diagrams, factors of $\pplus$ only enter through relativistic corrections and are therefore suppressed by the nucleon mass, which scales as \Nc. Since the analysis in the t-channel is sufficient to establish the large-\Nc scaling, momenta are counted as \cite{kaplan_nucleon-nucleon_1997}
    \begin{equation}
        \pminus \sim 1, \quad \pplus \sim \Nc^{-1}\, .
        \label{eq:p-scale}
    \end{equation}

Finally, large-\Nc scaling is impacted by the number of pions involved in the process.  In \chiPT, pion fields are encoded in the exponential matrix $u$ of Eq.~\eqref{eq:udef}. Expanding $u$ in the number of pions, we see that each pion field is accompanied by a factor of $1/F$.  In the large-\Nc limit the decay constant $F$ is $O(\sqrt{\Nc})$ \cite{t_hooft_planar_1974, witten_baryons_1979}; each additional pion field in the expansion of Eq.~\eqref{eq:udef} yields a suppression by $1/\sqrt{\Nc}$.  

In summary, the large-\Nc scaling of the LECs is determined by the spin-isospin structure of the matrix elements of nucleon bilinear operators, the scaling of any relevant momentum factors, and additional suppressions from any pion fields. Finally, the overall factor of \Nc in the Hamiltonian of Eq.~\eqref{Hartree} reduces the scaling of the LECs by one power of \Nc.

 This approach  has been used to analyze the large-\Nc behavior of \NN interactions in the symmetry-even \cite{kaplan_spin_1996,kaplan_nucleon-nucleon_1997,banerjee_nucleon_2002,Riska:2002vn,CalleCordon:2009ps,CalleCordon:2008cz,schindler_large-$n_c$_2018}  and symmetry-odd sectors \cite{schindler_large-$n_c$_2016, phillips_parity-violating_2014, samart_time-reversal_2016, vanasse_time-reversal-invariance_2019}, three-nucleon forces \cite{phillips_three-nucleon_2013}, and the coupling of two nucleons to external magnetic and axial fields \cite{richardson_large-_2020}. 
In this paper the large-\Nc scaling  is used to establish relationships among LECs associated with CIB $\NN$ operators.

We briefly comment on the role of the $\Delta$ in large-\Nc ChEFT.
The nucleon and $\Delta$ mass splitting is $O(1/\Nc)$; therefore, the nucleon and the $\Delta$ resonance become degenerate in the large-\Nc limit. 
The $\Delta$ was shown to play a crucial role in deriving the spin-flavor symmetry and obtaining consistent large-\Nc scaling for pion-baryon scattering \cite{Dashen:1993ac,Dashen:1993as,Gervais:1983wq,Gervais:1984rc}, as well as in understanding the meson-exchange picture of the \NN interactions \cite{banerjee_nucleon_2002}.
For the quantities of interest here, the $\Delta$ can only appear in intermediate states and effects of the virtual $\Delta$ degrees of freedom are not considered explicitly in the following. We thus obtain constraints on the LECs in a ChEFT that does not include explicit $\Delta$s.
Including the $\Delta$ resonance in \chiPT changes the size of the LECs and leads to quantities that may depend on the ratio 
    \begin{equation}
        \frac{m_\Delta - m_N}{ m_\pi },
    \end{equation}
which depends on the order in which the large-\Nc and chiral limits are taken \cite{Jenkins:1991ne,Cohen:1992uy,dashen_1/nc_1994,Cohen:1995mh}.
While this apparent difference in the treatment of the $\Delta$ between the large-\Nc and the EFT approaches is an important issue to be resolved, earlier work on the \NN interaction that similarly excluded intermediate $\Delta$ states obtained results that did not contradict available data \cite{kaplan_spin_1996,kaplan_nucleon-nucleon_1997,schindler_large-$n_c$_2018}. The role of intermediate $\Delta$ states in \NN scattering in the $\oneS$ channel and how they can be integrated out is discussed in Ref.~\cite{savage_delta_1997}.


\section{Chiral effective field theory and spurion fields}
    \label{Sec:Spurion}

The CIB ChEFT Lagrangian is constructed using the spurion technique, the same technique used to construct the mass term in the LO pion Lagrangian and to include the effects of virtual photons and leptons \cite{urech_virtual_1995, knecht_virtual_1998, neufeld_isospin_1995, neufeld_electromagnetic_1996, ecker_role_1989, knecht_chiral_2000, meisner_virtual_1997, meisner_isospin_1998, muller_virtual_1999}. The CIB Lagrangian of interest here contains terms with two insertions of the quark (or equivalently nucleon) charge matrix.
The QCD Lagrangian for two flavors in terms of left- and right-handed quark fields with minimal coupling to an electromagnetic potential is
    \begin{equation}\label{qL}
        \Lagr = i \bar q_L \slashed \partial q_L + i \bar q_R \slashed \partial q_R - \bar q_L M^\dagger q_R - \bar q_R M q_L + ie A_\mu \left[ \bar q_L \gamma^\mu Q_q q_L + \bar q_R \gamma^\mu Q_q q_R \right],
    \end{equation}
where $Q_q = \text{diag}(\frac{2}{3}, - \frac{1}{3})$ is the quark charge matrix and the unit of charge $e$ is factored out of $Q_q$ (see the last terms in Eq.~\eqref{qL}).
The quark mass and the charge matrix terms break chiral symmetry explicitly. For the mass terms, the pattern of symmetry breaking can be mapped onto the effective Lagrangian by ({\it i}) assuming that the constant matrix  $M$ transforms under the chiral symmetry group as
    \begin{equation}
        M \mapsto R M L^\dagger,
    \end{equation}
where $R$ and $L$ are SU(2) matrices transforming the right- and left-handed components of the quark fields, respectively, and ({\it ii}) constructing all allowed terms that are chirally invariant with the assumed transformation behavior of the quark mass matrix.
The same approach can be adopted for terms containing the charge matrix.  
First, $Q$ is separated into two matrices, $Q_L$ and $Q_R$, such that the electromagnetic part of the Lagrangian can be written as 
    \begin{equation}
        \Lagr_{EM} = ie A_\mu \left[ \bar q_L \gamma^\mu Q_L q_L + \bar q_R \gamma^\mu Q_R q_R \right] .
    \end{equation}
Next, the charge matrices are required to transform under the chiral symmetry group as
    \begin{align}
        Q_R &\mapsto R Q_R R^\dagger, \\
        Q_L &\mapsto L Q_L L^\dagger.
    \end{align}
At the nucleonic level, the Lagrangian can be written in terms of the nucleon doublet $N=(p,n)^T$, which transforms under chiral symmetry as
    \begin{equation}
        N \mapsto K(L,R,u) N, \quad K \in SU(2),
    \end{equation}
while the pion matrix
    \begin{equation}
        u \mapsto u^\prime = R u K^\dagger = K u L^\dagger \ .
    \end{equation} 
In Ref.~\cite{cirigliano_renormalized_2019}, the matrix containing the pion fields transforms as $U \mapsto L U R^\dagger$, where $u^2 = U$, while the transformation used here is $U \mapsto R U L^\dagger$ in accord with Ref.~\cite{gasser_chiral_1984}. However, this difference does not impact the results.
The construction of all possible nucleon operators with two spurion insertions that are invariant under chiral transformations is simplified by using the combinations
    \begin{equation}
        Q_{\pm} = \frac{1}{2} \left[ u^\dagger Q_R u \pm u Q_L u^\dagger \right],
        \label{Eq:Qplusminus}
    \end{equation}
which transform under the chiral symmetry group as 
    \begin{equation}
        Q_{\pm} \rightarrow K Q_{\pm} K^\dagger.
    \end{equation}
It is useful to separate these spurions into isoscalar and isovector components,
    \begin{equation}
        Q_\pm = \frac{1}{2} \Tr(Q_\pm) \1 + \tilde Q_\pm,
    \end{equation}
where 
    \begin{equation}
            \label{Traceless_isovector}
        \tilde Q_\pm = \frac{1}{2} \Tr(Q_\pm \tau^a) \tau^a,
    \end{equation}
so that the operators are written in terms of $\Tr(Q_\pm)$ and $\tilde Q_\pm$.


\section{Large-\Nc scaling of \NN interactions with two spurion fields}
    \label{Sec:LargeN_scaling}

The large-\Nc analysis discussed in Sec.~\ref{LNEFT} can be extended to include the spurion operators when an explicit form for the spurion is chosen. 
The greatest possible large-\Nc scaling of a given CIB operator can be deduced from its spin-flavor structure, which is used to guide the elimination of redundant operators  when Eq.~\eqref{Traceless_isovector} is inserted in the relevant nucleon bilinears (see Appendix~\ref{app:redundancy}).
However, as discussed above,  some operators may receive additional $1/\Nc$ suppressions when the leading term contains pion fields from the expansion of $u$. We will point out an explicit example of this in the next section.

One might attempt to obtain the large-\Nc scaling of \gnu, $\calC_1$, and $\calC_2$ directly from Eqs.~\eqref{CiriglianoLNV} and \eqref{CiriglianoCIB:1}. However, the forms of the Lagrangians in Eqs.~\eqref{CiriglianoLNV} and \eqref{CiriglianoCIB:1} are obtained by using Fierz identities to eliminate redundant operators. This procedure can obscure the correct large-\Nc scalings \cite{GirlandaCD15, schindler_large-$n_c$_2016}. Therefore, we present an alternative minimal basis in which the large-\Nc scaling of the LECs is manifest. 
The relationships between these LECs and the ones in Eqs.~\eqref{CiriglianoLNV} and \eqref{CiriglianoCIB:1} are given in Sec.~\ref{Sec:Justification}.

Instead of working in the basis of Ref.~\cite{cirigliano_neutrinoless_2017}, we will use the spurions defined in Eq.~\eqref{Eq:Qplusminus} and then translate between the two bases after the \LONc  Lagrangian has been derived.
When determining the large-\Nc scaling of general operator forms we will leave out  electromagnetic or weak factors such as $e^2$ or $\left(G_F V_{ud} \right)^2$. While these factors impact the overall size of an operator, they will not be relevant for understanding the relative large-\Nc rankings among operators that have the same overall multiplicative factor.
The most general set of operators for this analysis is given by
    \begin{eqnarray}
            \label{OperatorSets}
        B_1 & = & \Tr(Q_+)^2 \left( N^\dagger \Gamma N \right)^2 \, , \nonumber \\
        B_2 & = & \Tr(Q_+) \left( N^\dagger \Gamma N \right) \left( N^\dagger \tilde Q_+ \Gamma N \right)\, ,  \nonumber \\
        B_3 & = & \left( N^\dagger \tilde Q_+ \Gamma N \right)^2 \, , \nonumber \\
        B_4 & = & \Tr(Q_-)^2 \left( N^\dagger \Gamma N \right)^2 \, , \nonumber \\
        B_5 & = & \Tr(Q_-) \left( N^\dagger \Gamma N \right) \left( N^\dagger \tilde Q_- \Gamma N \right)\, ,  \nonumber \\
        B_6 & = & \left( N^\dagger \tilde Q_- \Gamma N \right)^2\, ,  \nonumber \\
        B_7 & = & \Tr(Q_+) \Tr(Q_-) \left( N^\dagger \Gamma N \right) \left( N^\dagger \Gamma N \right) \, ,  \nonumber \\
        B_8 & = & \Tr(Q_-) \left( N^\dagger \tilde Q_+ \Gamma N \right) \left( N^\dagger \Gamma N \right) \, ,  \nonumber \\
        B_9 & = & \Tr(Q_+) \left( N^\dagger \tilde Q_- \Gamma N \right) \left( N^\dagger \Gamma N \right) \, ,\nonumber \\
        B_{10} & = & \left( N^\dagger \tilde Q_+ \Gamma N \right) \left( N^\dagger \tilde Q_- \Gamma N \right) \, , \nonumber \\
        B_{11} & = & \Tr(\tilde Q_+^2 + \tilde Q_-^2) \left( N^\dagger \Gamma N \right)^2 = \frac{1}{2} \Tr(\tilde Q_R^2 + \tilde Q_L^2) \left( N^\dagger \Gamma N \right)^2 \, ,\nonumber \\
        B_{12} & = & \Tr(\tilde Q_+^2 - \tilde Q_-^2) \left( N^\dagger \Gamma N \right)^2 = \Tr( U \tilde Q_L U^\dagger \tilde Q_R ) \left( N^\dagger \Gamma N \right)^2 \, , \nonumber \\
        B_{13} & = & \Tr(\tilde Q_+ \tilde Q_-) \left( N^\dagger \Gamma N \right)^2 = \Tr(\tilde Q_R^2 - \tilde Q_L^2) \left( N^\dagger \Gamma N \right)^2 \, ,
    \end{eqnarray}
where $\Gamma$ can be $\1$, $\sigma^i$, $\tau^a$, or $\sigma^i \tau^a$. However, several of the  operators that arise once all four of the possibilities for $\Gamma$ are inserted into the general forms of Eq.~\eqref{OperatorSets} will be redundant.
The nucleon bilinears contained in operators from $B_1$, $B_4$, $B_7$, $B_{11}$, $B_{12}$, and $B_{13}$ have the same structure as the operators from nucleon-nucleon scattering, so  operators with $\Gamma = \1$ and $\sigma^i \tau^a$ may start to contribute at LO in \Nc, while those with $\Gamma = \sigma^i$ and $\tau^a$ are $1/\Nc^2$ suppressed. The Fierz identity
    \begin{equation}
        \label{eq:NNFierz1} 
        \left( N^\dagger \sigma^i \tau^a N\right)^2 = -3 \left( N^\dagger N\right)^2
    \end{equation} 
can be used to eliminate $\sigma^i \tau^a$ in favor of $\1$, and since the corresponding LECs are of the same order there is no change in the 
scaling obtained. Similarly, the identity 
    \begin{equation}
        \label{eq:NNFierz2}
        \left( N^\dagger \tau^a N\right)^2  = -2 \left( N^\dagger N \right)^2 - \left( N^\dagger \sigma^i N \right)^2
    \end{equation}
shows that the bilinear with $\tau^a$  is not independent of those containing $\1$ and $\sigma^i$ in the operators of the form $B_1$, $B_4$, $B_7$, $B_{11}$, $B_{12}$, and $B_{13}$.

For operators of the form $B_2$, $B_3$, $B_5$, $B_6$, $B_8$, $B_9$, and $B_{10}$, on the other hand, the insertion of $\Gamma = \tau^a$ or $\sigma^i \tau^a$ creates terms containing products of Pauli matrices in isospin space in a single nucleon bilinear. 
The structure of these terms does not match directly onto the Hartree Hamiltonian of Eq.~\eqref{Hartree}. But the terms can be rewritten using
   \begin{equation}
             \label{PauliProduct}
         \tau^a \tau^b = \delta^{ab} \1 + i \epsilon^{abc} \tau^c \ , 
     \end{equation}
which generates structures that contain at most a single isospin Pauli matrix. Again, the large-\Nc scaling of these terms can be determined from Eq.~\eqref{MEScaling}, and the forms with $\Gamma = \tau^a$ and $\sigma^i \tau^a$ can be eliminated for this set of operators. 
There is one more redundancy. Operators with $\Gamma = \1$ or $\sigma^i$ can be removed through the use of Eq.~\eqref{Traceless_isovector} and Fierz transformations. For $B_3$, $B_6$, and $B_{10}$, $\Gamma = \1$ can be eliminated, while either choice is suitable for $B_2$, $B_5$, $B_8$, and $B_9$ since both choices scale with \Nc in the same way. Again, Appendix \ref{app:redundancy} contains greater detail about this procedure. In the next two sections, the explicit forms of the spurion fields for the electromagnetic and the weak cases, respectively, are considered.


\subsection{Electromagnetic Spurions}
\label{sec:emspurions}

For the electromagnetic case, it is useful to write the Lagrangian in terms of the nucleon charge matrix,
    \begin{equation}\label{Q}
        Q = \frac{1}{2} \left( \1 + \tau^3 \right).
    \end{equation}
The difference between using the nucleon charge matrix and using the quark charge matrix amounts to a shift by an unobservable constant \cite{meisner_isospin_1998}. 
Here, the nucleon charge matrix is independent of \Nc, which implies that the up and down quark charges are \Nc-dependent. The alternative choice that the quark charges are constant and the nucleon charge depends on \Nc is discussed in Appendix~\ref{app:charge}.

Using Eq.~\eqref{Q} as the spurion field in Eq.~\eqref{OperatorSets} yields operators with a clear spin-flavor structure. Setting $Q_R = Q_L = Q$ gives
    \begin{eqnarray}
        \tilde Q_\pm & = & \frac{1}{4} \left[ u^\dagger \tau^3 u \pm u \tau^3 u^\dagger \right].
    \end{eqnarray}
The corresponding traces of operators are \cite{muller_virtual_1999}
    \begin{eqnarray}
        \Tr(Q_+) & = & 1, \\
        \Tr(Q_-) & = & 0, \\
        \Tr(\tilde Q_+^2 + \tilde Q_-^2) & = & \Tr(\tilde Q^2) = 1/2, \\
        \Tr(\tilde Q_+^2 - \tilde Q_-^2) & = & \Tr(U \tilde Q U^\dagger \tilde Q), \\
        \Tr(\tilde Q_+ \tilde Q_-) & = & 0.
    \end{eqnarray}
Since 
    \begin{eqnarray}
        \Tr( \tilde Q_+^2 + \tilde Q_-^2 ) \left( N^\dagger \calO N \right)^2 & \sim & \Tr(Q_+)^2 \left( N^\dagger \calO N \right)^2\ ,
    \end{eqnarray}
the operators from $B_{11}$ can be absorbed into those from $B_1$.
Therefore, the only independent operators are those from $B_1$, $B_2$, $B_3$, $B_6$, $B_9$, $B_{10}$, and $B_{12}$. 
The operators from $B_{10}$ vanish at least through $O(\phi^2)$ when $u$ is expanded and can be neglected at this order.
After eliminating redundancies (see Appendix~\ref{app:redundancy}) the remaining operators are
    \begin{eqnarray}
        \calO_{1,1} & = & \left( N^\dagger N \right)^2 \, , \\
        \calO_{1,2} & = & \left( N^\dagger \sigma^i N \right)^2 \, , \\
        \calO_{2} & = & \left( N^\dagger N \right) \left( N^\dagger \tilde Q_+ N \right) \, , \\
        \calO_3 & = & \left( N^\dagger \sigma^i \tilde Q_+ N \right) \left( N^\dagger \sigma^i \tilde Q_+ N \right) \, , \\
        \calO_6 & = & \left( N^\dagger \sigma^i \tilde Q_- N \right) \left( N^\dagger \sigma^i \tilde Q_- N \right) \, , \\
        \calO_9 & = & \left( N^\dagger \tilde Q_- N \right) \left( N^\dagger N \right) \, , \\
        \calO_{12,1} & = & \Tr( U \tilde Q U^\dagger \tilde Q ) \left( N^\dagger N \right)^2 \, , \\
        \calO_{12,2} & = & \Tr( U \tilde Q U^\dagger \tilde Q ) \left( N^\dagger \sigma^i N \right)^2  \, ,
    \end{eqnarray}
where the first subscript $i$ in $\calO_{i,j}$ indicates the $B_i$ from which each operator originates, and the second index $j$, where necessary,  refers to a specific operator within the $B_i$, $j=1,2,3,4$ for $\Gamma$= $\1$, $\sigma^i$, $\tau^a$, $\sigma^i \tau^a$, respectively.

Finally, as discussed in Sec.~\ref{LNEFT}, each additional pion field introduces a factor of $1/F\sim 1/\sqrt{\Nc}$.
Expanding  each operator to second order in the pion fields to determine the maximum \Nc scaling of the corresponding LECs yields 
\begin{align}
        \calO_{1,1} & =  \left( N^\dagger N \right)^2 + \cdots \label{eq:O11} \, ,\\
        \calO_{1,2} & =  \left( N^\dagger \sigma^i N \right)^2 + \cdots  \, ,\\
        \calO_{2} & =  \frac{1}{2} \left( 1 - \frac{1}{2 F^2} \phi_a \phi_a \right) \left( N^\dagger N \right) \left( N^\dagger \tau^3 N \right) + \frac{1}{4 F^2} \phi_3 \phi_a \left( N^\dagger N \right) \left( N^\dagger \tau^a N \right)  + \cdots \, , \\
        \calO_3 & =  \frac{1}{4}  \left( 1 - \frac{1}{F^2} \phi_a \phi_a \right) \left( N^\dagger \sigma^i \tau^3 N \right)^2 + \frac{1}{4 F^2} \phi_3 \phi_a  \left( N^\dagger \sigma^i \tau^3 N \right) \left( N^\dagger \sigma^i \tau^a N \right) + \cdots \, , \\
        \calO_6 & =  \frac{1}{4 F^2} \epsilon^{3ab} \epsilon^{3cd} \phi_a \phi_c \left( N^\dagger \sigma^i \tau^b N \right) \left( N^\dagger \sigma^i \tau^d N \right) + \cdots  \, ,\\
        \calO_9 & =  - \frac{1}{2 F} \epsilon^{3ab} \phi_a \left( N^\dagger \tau^b N \right) \left( N^\dagger N \right) + \cdots  \, ,\\
        \calO_{12,1} & =   \left[ \frac{1}{2} - \frac{4}{F^2} \left( \phi_a \phi_a - \phi_3 \phi_3 \right)  \right] \left( N^\dagger N \right)^2 + \cdots  \, , \\
        \calO_{12,2} & =  \left[ \frac{1}{2} - \frac{4}{F^2} \left( \phi_a \phi_a - \phi_3 \phi_3 \right)  \right] \left( N^\dagger \sigma^i N \right)^2 + \cdots \,  \label{eq:O122}
    \end{align}
where the ellipses indicate additional pion fields. The scaling of the LECs $\bar {\cal C}_{i,j}$ multiplying $\calO_{i,j}$ in the Lagrange density is given by
    \begin{align}
        \bar \calC_{1,1}&\sim \Nc \label{eq:C11} \, ,\\
        \bar \calC_{1,2} & \sim \Nc^{-1}  \, ,\\
        \bar \calC_2 & \sim 1 \, , \\
        \bar \calC_3 &\sim \Nc \, , \\
        \bar \calC_{6} &\sim 1  \, ,\\
        \bar \calC_{9} &\sim \Nc^{-1/2}  \, ,\\
        \bar \calC_{12,1} &\sim \Nc  \, , \\
        \bar \calC_{12,2} &\sim \Nc^{-1} \, . \label{eq:C122}
    \end{align}
The operators $\calO_{1,1}$ and $\calO_{12,1}$ differ only at the multi-pion level. Therefore, differences between the two will be 1/\Nc suppressed. 
The same holds for the operators $\calO_{1,2}$ and $\calO_{12,2}$.
The operator $\calO_6$ provides a concrete example of an earlier point: the generic spin-flavor structure of the operator, before expanding $u$ in the number of pion fields, indicates that it could be $O(\Nc$), but the first nonzero term has two pion fields and is thus suppressed by an additional factor of $1/\Nc$.

The Lagrangian at LO and next-to-leading order (NLO) in the large-\Nc expansion  is
    \begin{align}
        \Lagr_{\LONc} & =  e^2\left\{\left[ \bar \calC_{1,1} + \bar \calC_{12,1} \Tr(U \tilde Q U^\dagger \tilde Q) \right] \left( N^\dagger N \right)^2 + \bar \calC_{3} \left( N^\dagger \sigma^i \tilde Q_+ N \right)^2\right\} \, , \label{eq:LONc:1} \\
        \Lagr_{\NLONc} & =  e^2\left\{\bar \calC_{2} \left( N^\dagger N \right) \left( N^\dagger \tilde Q_+ N \right) + \bar \calC_{6} \left( N^\dagger \sigma^i \tilde Q_- N \right)^2 \right\}\label{eq:NLONc:1} \, .
    \end{align}
The $\bar \calC_{i}$ and $\bar \calC_{i,j}$ are LECs that have to be determined from comparison to data or from a calculation in terms of the underlying QCD degrees of freedom.
Expanding the matrices $u$ and $U$ in the number of pion fields also creates terms at higher order in the large-\Nc counting than indicated by the subscript on the left side; see the discussion in Sec.~\ref{LNEFT}.
In Sec.~\ref{Sec:Justification}, we will map  the form of the Lagrangian in Eqs.~\eqref{eq:LONc:1} and \eqref{eq:NLONc:1} to the one used in Eq.~\eqref{CiriglianoCIB:1} to determine the large-\Nc scaling of the LECs in Eq.~\eqref{CiriglianoCIB:1}.


\subsection{Weak Spurions}
\label{sec:weakspurions}
For weak interactions, $Q_L$ is given by Eq.~\eqref{eq:weakspurion} while $Q_R = 0$, which gives
\begin{equation}
    Q_\pm  = \tilde Q_\pm  = \pm \frac{1}{2} u Q_L u^\dagger = \pm \frac{1}{2} u \tau^+ u^\dagger \ .
\end{equation}
As a result, all traces in Eqs.~\eqref{OperatorSets} vanish and therefore operators from $B_1$, $B_2$, $B_4$, $B_5$, $B_7$, $B_8$, $B_9$, $B_{11}$, $B_{12}$, and $B_{13}$ do not contribute. Since $\tilde Q_+ = - \tilde Q_-$, the only nonvanishing term is
    \begin{equation}
        \left( N^\dagger u \tau^+ u^\dagger \Gamma N \right)^2,
    \end{equation}
and the structures $B_3$, $B_6$, and $B_{10}$ become identical. As pointed out in Ref.~\cite{cirigliano_neutrinoless_2017}, the two operators corresponding to $\Gamma = \1$ and $\sigma^i$ in this term  are related through a Fierz identity and are not independent at $O(\phi^0)$. The authors of Ref.~\cite{cirigliano_neutrinoless_2017} choose to retain $\Gamma = \1$; that is, the operator $\left( N^\dagger \tau^+ N \right)^2$. According to Eq.~\eqref{MEScaling}, this operator does not appear at LO  in the large-\Nc expansion. However, eliminating the operator $\left( N^\dagger \sigma^i \tau^+ N \right)^2$ through the Fierz transformation
    \begin{equation}
        \left( N^\dagger \sigma^i \tau^+ N \right)^2 = - 3 \left( N^\dagger \tau^+ N \right)^2 
    \end{equation}
introduces a hidden \LONc contribution in the term proportional to $\left( N^\dagger \tau^+ N \right)^2$. As a result, after removing the overall factor of $\Nc$ from the Hartree Hamiltonian as discussed in Sec.~\ref{LNEFT}, \gnu is of LO in the large-\Nc expansion, $\gnu \sim O(\Nc)$. 
This result by itself does not justify the assumptions underlying the approximation $\gnu \approx \frac{1}{2}(\calC_1+\calC_2)$  proposed in Refs.~\cite{cirigliano_new_2018,cirigliano_renormalized_2019}. However, an inconsistency in the large-\Nc scaling of \gnu versus $(\calC_1+\calC_2)$  would cast doubt on the approximation. As will be shown in Sec.~\ref{Sec:Justification}, $(\calC_1+\calC_2)\sim O(\Nc)$, consistent with the LO scaling of \gnu found here.


\section{Large-\Nc hierarchy of charge-independence-breaking interactions}
    \label{Sec:hierarchy}

Before focusing on the isotensor terms and their relation to $\gnu$, we will analyze the large-\Nc scaling of general CIB \NN interactions using the results of Sec.~\ref{sec:emspurions}.
In the absence of external pions, the operators in Eqs.~\eqref{eq:O11} - \eqref{eq:O122} that contain pions only contribute through pion-loop diagrams that are of higher order in the chiral power counting than is considered in this analysis. 
Adopting the conventions in  Ref.~\cite{Henley:1977qc} (also see Ref.~\cite{miller_charge_2006}), the \NN interactions, including the effects of virtual photons, are divided into four classes characterized by the following  isospin  structures:
\begin{enumerate}
    \item[(I)] isospin invariant and charge symmetric: $\1_1\1_2$, $\vec\tau_1 \cdot \vec\tau_2$,
    \item[(II)] CIB but not charge-symmetry-breaking (CSB), which have the isotensor form: $\tau^3_1\tau^3_2 - \frac{1}{3} \vec\tau_1 \cdot \vec\tau_2 $,
    \item[(III)] CSB (and thus CIB) terms that are symmetric in spin and isospin indices: $\tau^3_1 + \tau^3_2$,
    \item[(IV)] CSB with isospin mixing (these vanish on $nn$ and $pp$ systems, but not $np$, and only occur in $L \neq 0$ partial waves): $\tau^3_1 -\tau^3_2$, $(\vec\tau_1 \times \vec\tau_2)^3$.
\end{enumerate}
The subscripts in the expressions above denote nucleon bilinears one and two. Refs.~\cite{van_kolck_soft_1993, miller_charge_2006, miller_charge_1994} use dimensional analysis to argue that  the size of these interactions is such that Class (I) $>$  Class (II) $>$ Class (III) $>$ Class (IV).

Neglecting the operators $\calO_6$ and $\calO_9$ because they contain at least one pion field, the independent contact operators generated by the spurion formalism fall into the categories
    \begin{eqnarray}
        \text{(I)} & \ & \calO_{1,1}, \ \calO_{1,2} \ ,\\
        \text{(II)} & \ & \calO_{3} \ ,\\
        \text{(III)} & \ & \calO_{2} \, .
    \end{eqnarray}
As discussed in Sec.~\ref{sec:emspurions}, the pionless parts of the operators $\calO_{12,1}$ and $\calO_{12,2}$ are identical to $\calO_{1,1}$ and $\calO_{1,2}$, respectively. Class (I) and (II) interactions appear at the same order in the large-\Nc expansion, while Class (III) terms are suppressed by $1/\Nc$. 
It may be unexpected that the large-\Nc analysis suggests that the isospin-invariant Class (I) interactions appear at the same order as CIB terms. Recall, though, that the operators considered here are accompanied by factors of $e^2$ in the Lagrangian. The Class (I) terms derived here are therefore  $O(e^2)$-suppressed corrections to the dominant isospin-invariant interactions. 
Taking into account the additional $e^2$ suppression of the isospin-violating terms, our results are not in contradiction with the expectations of Refs.~\cite{van_kolck_soft_1993, miller_charge_2006, miller_charge_1994} that some Class (I) terms are larger than Class (II) terms.
Contact operators leading to Class (IV) CIB contain two derivatives and are of higher order in the EFT expansion. Taking into account the scaling of the momenta in Eq.~\eqref{eq:p-scale}, these terms are at most $O(\Nc^0)$. 
Additionally, at the level of the \NN Lagrangian, the two operators that lead to the Class (IV) potential given in \cite{miller_charge_2006} are related by Fierz transformations and are not independent at the two-derivative order in the EFT expansion. 
But previous work \cite{Lynn:2015jua,Lynn:2017fxg,Lonardoni:2018nob} has shown that formally Fierz-equivalent operators can lead to ambiguities when used in deriving potentials with local regulators.


\section{Large-\Nc justification for $\gnu \approx \frac{1}{2}(C_1+C_2)$~\cite{cirigliano_renormalized_2019}}
    \label{Sec:Justification}

To connect Eq.~\eqref{eq:LONc:1} to Eq.~\eqref{CiriglianoCIB:1}, it is helpful to rearrange the \LONc Lagrangian (Eq.~\eqref{eq:LONc:1}) as
    \begin{equation}
        \begin{split}
        \Lagr_{\LONc}  = & e^2  \left\{ \frac{1}{2} \left[ 2 \bar \calC_{1,1} + \bar \calC_{12,1} - \bar \calC_{3} \right] \Tr( \tilde Q_+^2 ) \left( N^\dagger N \right)^2 \right. \\
        & + \left. \bar \calC_{3} \left[ \left(N^\dagger \sigma^i \tilde Q_+ N \right)^2 - \frac{1}{6} \Tr(\tilde Q_+^2) \left( N^\dagger \sigma^i \tau^a N \right)^2 \right] \right\}, 
        \end{split}
    \end{equation}   
where the second term proportional to $\bar \calC_{3}$ is now a symmetric traceless isotensor. The included trace term appears at the same order in the large-\Nc expansion.
This rearrangement also produces a \NLONc contribution such that Eq.~\eqref{eq:NLONc:1} becomes
    \begin{equation}
            \label{eq:NLONc:2}
    \begin{split}
        \Lagr_{\NLONc}  =  & \, e^2 \left\{ \frac{1}{2} \left[ 2 \bar \calC_{1,1} - \bar \calC_{12,1} - \bar \calC_{6} \right] \Tr( \tilde Q_-^2 ) \left( N^\dagger N\right)^2  + \bar \calC_{2} \left( N^\dagger N \right) \left( N^\dagger \tilde Q_+ N \right) \right.  \\
        & \left. + \, \bar \calC_{6} \left[ \left( N^\dagger \sigma^i \tilde Q_- N \right)^2 - \frac{1}{6} \Tr( \tilde Q_-^2 ) \left( N^\dagger \sigma^i \tau^a N \right)^2 \right] \right\}.
    \end{split}
    \end{equation}
We now consider the isotensor CIB term proportional to $\bar \calC_{3}$ in more detail, and relate it to the terms used in Ref.~\cite{cirigliano_renormalized_2019}, see Eq.~\eqref{CiriglianoCIB:1}. 
Fierz transformations are used to rewrite the leading terms (see Eq.~\eqref{eq:traceless_tensor_fierz}). This uncovers \LONc scaling in terms that naively appear to be of higher order. The resulting Lagrangian is 
    \begin{equation}
            \label{eq:PostFierz}
        \Lagr_{\LONc}^{\Delta I = 2}  =  -3 e^2 
         \bar \calC_{3} \left[ \left(N^\dagger \tilde Q_+ N \right)^2 - \frac{1}{6} \Tr(\tilde Q_+^2) \left( N^\dagger \tau^a N \right)^2 \right].
    \end{equation}

Using the definition of the spurion fields in Eqs.~\eqref{Eq:Qplusminus}, the Lagrangian of Eq.~\eqref{CiriglianoCIB:1} can be written as
    \begin{eqnarray}
          \label{CirgilianoCIB:Translation}
        \Lagr_{CIB}^{\NN} & = & \frac{e^2}{2} \left\{ \left( \calC_1 + \calC_2 \right) \left[ \left( N^\dagger \tilde Q_+ N \right)^2 - \frac{1}{6} \Tr( \tilde Q_+^2) \left( N^\dagger \tau^a N \right)^2  \right] \right. \nonumber \\
        & & + \left. \left( \calC_1 - \calC_2 \right) \left[ \left( N^\dagger \tilde Q_- N \right)^2 - \frac{1}{6} \Tr(\tilde Q_-^2) \left( N^\dagger \tau^a N \right)^2 \right] \right\} \ .
    \end{eqnarray}
Comparison with Eq.~\eqref{eq:PostFierz} shows that 
    \begin{equation}
       \frac{1}{2}\left( \calC_1 + \calC_2 \right) = -3 \bar \calC_{3} ,
    \end{equation}
which demonstrates that $\calC_1 + \calC_2 \sim \Nc$. 
A similar transformation for the isotensor term in Eq.~\eqref{eq:NLONc:2} shows that 
    \begin{equation}
        \frac{1}{2} \left( \calC_1 - \calC_2 \right) = -3 \bar \calC_{6} \ ,
    \end{equation}
demonstrating that $\calC_1 - \calC_2$ is 1/\Nc suppressed relative to $\calC_1 + \calC_2$.
Inverting these equations gives
\begin{align}
    \calC_1 & =  -3\bar\calC_3 - 3\bar\calC_6 = -3\bar\calC_3 \left[ 1+O(1/\Nc)\right] , \\
    \calC_2 & = -3\bar\calC_3 + 3\bar\calC_6 = -3\bar\calC_3 \left[ 1+O(1/\Nc)\right].
\end{align}
These results support the assumption of Ref.\cite{cirigliano_renormalized_2019} that the LECs in the CIB Lagrangian are of the same size and sign,  and that therefore the neutrinoless LEC can be approximated as $\gnu \approx\frac{1}{2}(\calC_1 + \calC_2)$.


\section{Conclusion}
    \label{Sec:conclusion}

The renormalization group analysis of Refs.~\cite{cirigliano_new_2018,cirigliano_renormalized_2019} showed that, for light-Majorana exchange, an LNV contact term is required at leading order in ChEFT. The presence of this term impacts the calculation of nuclear matrix elements relevant for \neutrinoless decay. Neither sufficient data nor lattice QCD results are currently available to determine the size of the corresponding LEC, \gnu. 
To estimate the contribution of the LNV contact term to nuclear matrix elements, Refs.~\cite{cirigliano_new_2018,cirigliano_renormalized_2019} assumed that the two CIB LECs $\calC_1$ and $\calC_2$ are of the same size and sign, which allowed them to approximate $\gnu\approx (\calC_1 +\calC_2)/2$.

Here, we performed large-\Nc analyses of the LNV and CIB \NN operators appearing at the first nonvanishing order in ChEFT power counting.
Our results show that the assumptions underlying the approximations of \gnu used in Refs.~\cite{cirigliano_new_2018,cirigliano_renormalized_2019} are consistent with ordering based upon the large-\Nc limit, lending additional support to the numerical estimates for matrix elements found there.
They are also in line with the recent results of Refs.~\cite{Cirigliano:2020dmx, Cirigliano:2021qko}.

Our analysis also shows a hierarchy of the different classes of CIB \NN interactions as defined in Refs.~\cite{Henley:1977qc}. The ordering obtained does not contradict phenomenological expectations \cite{van_kolck_soft_1993, miller_charge_2006, miller_charge_1994}.
However, as is generally the case, the large-\Nc results should not be treated as precise predictions. The ordering of LECs  is based on expansions in 1/\Nc and the assumption that other numerical factors are of natural size. For example, symmetries not captured by the large-\Nc expansion may lead to unnaturally small parameters.
In particular, lattice QCD calculations of baryon-baryon interactions suggest that there is an accidental SU(16) symmetry beyond the SU(6) symmetry in three-flavor large-\Nc QCD \cite{Wagman:2017tmp, Illa:2020nsi}.
Two additional caveats to the results in this paper are that there are unresolved open questions involving the application of large-\Nc scaling of operators within heavy nuclei, and the potential impact of $\Delta$ intermediate states. So far these issues have not exposed any practical flaws to the procedure used in this paper, but they should be kept in mind.  We hope that this work will help guide many-body studies of LNV in heavier elements, as well as the interpretation of neutrinoless double beta decay experiments.


\begin{acknowledgments}

We thank Emanuele Mereghetti for useful discussions.  
We thank the Institute for Nuclear Theory at the University of Washington for its kind hospitality and stimulating research environment during the INT-18-2a program ``Fundamental Physics with Electroweak Probes of Light Nuclei.'' This research was supported in part by the INT's U.S.~Department of Energy grant No.~DE-FG02-00ER41132.
This material is based upon work supported by the U.S.~Department of Energy, Office of Science, Office of Nuclear Physics,
under Award Numbers DE-SC0019647 (T.R.R. and M.R.S.), DE-SC0021027 (S.P.), and DE-FG02-05ER41368 (R.P.S.). 

\end{acknowledgments}


\appendix

\section{Alternative Electric Charge Scaling}
\label{app:charge}

The choice of keeping the nucleon charge independent of \Nc, while the quark charges scale with \Nc, has the advantage that anomaly cancellations persist in a large-\Nc extended standard model  \cite{shrock_implications_1996, chow_selfconsistent_1996}. The up and down quark charges in units of $e$  are then given by 
    \begin{equation}
            \label{QuarkCharge}
        q_u = \frac{\Nc + 1}{2 \Nc}, \ q_d = \frac{1 - \Nc}{2 \Nc} \, ,
    \end{equation}
where \Nc is odd but arbitrary. 
This choice  leads to a proton with electric charge of one in units of $e$ when it is taken to consist of $\frac{1}{2} \left( \Nc + 1 \right)$ up quarks and $\frac{1}{2} \left( \Nc - 1 \right)$ down quarks. 
Similarly, the neutron has electric charge 0 when the numbers of quark flavors are switched.

In the meson sector of \chiPT, it is customary to use the quark charge matrix when constructing the spurion counterterms. However, when nucleons are included it is conventional to use the nucleon charge matrix. The terms in the pion Lagrangian are then replaced accordingly, but this only amounts to the addition of an unobservable constant term. When going to large-\Nc, it is reasonable to ask if this is still the case when the charge matrices with different large-\Nc scalings are interchanged. To answer this question,  the quark and nucleon charge matrices are generalized \cite{muller_virtual_1999}, 
    \begin{equation}
        Q = \alpha \1 + \beta \tau^3 \, .
    \end{equation}
The leading order operator in the pion Lagrangian is 
    \begin{equation}
        e^2 C \Tr(Q U Q U^\dagger) = e^2 C \Tr(\alpha^2 \1 + \beta^2 \tau^3 U \tau^3 U^\dagger).
    \end{equation}
The first term is indeed an unobservable constant shift, and the second term leads to the electromagnetic pion mass splitting when $U$ is expanded to $O(\phi^2)$, i.e.
    \begin{equation}
            \label{PionMass}
        \delta m_\pi^2 = \frac{2 e^2}{F_0^2} C  \, .
    \end{equation}

For quark charges that scale as Eq.~(\ref{QuarkCharge}), the quark and nucleon charge matrices become 
\begin{eqnarray}
       Q_{q}^\text{q-scaling} & = & \frac{1}{2 \Nc} \left[ \1 + \Nc \tau^3 \right] \, , \\
       Q_{N}^\text{q-scaling} & = & \frac{1}{2} \left[ \1 + \tau^3 \right] \, ,
    \end{eqnarray}
where the superscript indicates that the quark $q=u,d$ charges scale with $\Nc$.

Alternatively, it was argued that for baryons containing $O(\Nc^0)$ strange quarks, quantization conditions require the quark charges be fixed to their physical values and independent of \Nc \cite{Cohen:2003mc}. However, for this choice, anomalies in an SU(\Nc)-extended standard model do not cancel \cite{chow_selfconsistent_1996,shrock_implications_1996} and the nucleon charge becomes \Nc -dependent and unbounded as $\Nc\to\infty$. 
Nevertheless, as shown in the following, such a choice does not change our conclusions.
The quark and nucleon charge matrices are then 
    \begin{eqnarray}
        Q_{q}^{\text{q-fixed}} & = & \frac{1}{6} \left[ \1 + 3 \tau^3 \right] \, , \\
        Q_{N}^{\text{q-fixed}} & = & \frac{1}{6} \left[ \Nc \1 + 3 \tau^3 \right] \,
        \label{eq:scaling_nucleon_charge},
    \end{eqnarray}
where the superscript indicates that the quark charges $q$ are fixed as $\Nc$ changes. 
Regardless of whether the quark or nucleon charge matrices are chosen to scale with \Nc, the coefficient $\beta = \frac{1}{2}$. Therefore, both choices lead to the same pion mass splitting.

\begin{figure}[h]
    \centering
    \includegraphics{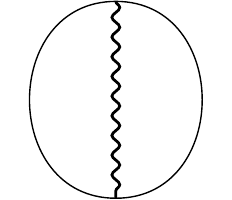}
    \caption{Leading order diagram in large-\Nc QCD, which is $O(e^2 \Nc)$.}
    \label{fig:quark_loop}
\end{figure}

Based on the argument that a single flavor trace operator in the meson sector of \chiPT corresponds to a single closed loop in large-\Nc QCD, it might be expected that the LEC $C$ scales at most as $C \sim \Nc$.
Using the typical diagrammatic arguments in Fig.~\ref{fig:quark_loop},  adding a photon in the loop does not modify the color structure, so it still consists of a single sum over all colors but it does pick up a factor of $e^2$. Therefore, the pion mass splitting in Eq.~(\ref{PionMass}) is at most $O(\Nc^0)$ when  $e$ is taken to be fixed and after accounting for the suppression due to $F_0$. 
However, it was shown (see, e.g., Ref. \cite{shrock_implications_1996,chow_selfconsistent_1996}) that for electroweak effects to be finite, the electromagnetic coupling can be rescaled like the strong coupling, i.e. $e \sim \Nc^{-1/2}$. 
In this case, the mass splitting will be $O(1/\Nc)$. 

When the nucleon charge is chosen to have the $\Nc$ dependence given by Eq.~\eqref{eq:scaling_nucleon_charge}, the large-\Nc behavior of the operators in Eqs.~(\ref{OperatorSets}) needs to be reexamined for possible changes. The operators that contain $\Tr(Q_\pm)$ are now multiplied by an overall factor of \Nc for each insertion of the trace. This leads to \
    \begin{align}
        \bar\calC_{1,1} & \sim \Nc^{3} \, , \\
        \bar\calC_{1,2} & \sim \Nc  \, ,\\
        \bar\calC_{2} & \sim \Nc  \, ,\\
        \bar\calC_{9} &\sim \Nc^{1/2} \, , 
    \end{align}
while the large-\Nc scaling of the other LECs remains unchanged.
The operators relevant for the classification of the CIB terms are still $\calO_{1,1}$, $\calO_{1,2}$, $\calO_{2}$, and $\calO_{3}$. To obtain the traceless form of the Class (II) interactions, the term in the Lagrangian containing $\calO_{3}$ can be rewritten as
\begin{equation}
    (N^\dagger \sigma^i \tau^3 N)^2 \\
    = (N^\dagger \sigma^i \tau^3 N)^2 - \frac{1}{3} (N^\dagger \sigma^i \tau^a N)^2 + \frac{1}{3} (N^\dagger \sigma^i \tau^a N)^2 \, .
\end{equation}
The first two terms on the right-hand side combine to form the Class (II) interaction. The last term is absorbed as an $O(\Nc)$ contribution into the Class (I) interaction.
With the alternative large-\Nc scaling of the charges, the classes of charge dependence are then altered such that (II) and (III) are the same order in \Nc while they are both suppressed by $\Nc^{-2}$ relative to (I).
This also indicates that the correspondence between the LNV operator and the CIB contact term remains intact regardless of the choice taken for the scaling of the nucleon charge with $\Nc$.


\section{Fierz identities and the elimination of redundant operators}
    \label{app:redundancy}    
The operators that contain only traces of the spurions have the form $(N^\dagger \Gamma N)^2$, where $\Gamma$ can be $\1$, $\sigma^i$, $\tau^a$, or $\sigma^i \tau^a$. The Fierz identities in Eqs.~\eqref{eq:NNFierz1} and \eqref{eq:NNFierz2} from Sec.~\ref{Sec:LargeN_scaling},
\begin{align*}
    \left( N^\dagger \sigma^i \tau^a N\right)^2 & = -3 \left( N^\dagger N\right)^2 \, , \tag{\ref{eq:NNFierz1}}\\
    \left( N^\dagger \tau^a N\right)^2  & = -2 \left( N^\dagger N \right)^2 - \left( N^\dagger \sigma^i N \right)^2 \, , \tag{\ref{eq:NNFierz2}}
\end{align*}
reduce the number of independent operators from four to two;
    \begin{eqnarray}
        \left( N^\dagger N \right)^2 \, , \\
        \left( N^\dagger \sigma^i N \right)^2 \, ,    
    \end{eqnarray}
where the first operator is \LONc, and the second is $1/\Nc^2$ suppressed.

For the operators involving the traceless part of the spurion field in the bilinears, the spurions in Eq.~\eqref{Traceless_isovector} are expanded and the products of Pauli matrices reduced using  Eq.~\eqref{PauliProduct}.
All of the suppressions arising from the presence of pion fields are contained in  coefficients defined by $c_{a, \pm} = \frac{1}{2} \Tr(\tilde Q_\pm \tau^a)$.
Therefore, $\Gamma = 1$ and $\sigma^i$, respectively, lead to
    \begin{eqnarray}
        \left(N^\dagger \tilde Q_\pm  N \right) \left( N^\dagger \tilde Q_\pm  N \right) & = & c_{a, \pm} c_{b, \pm} \left( N^\dagger \tau^a N \right) \left( N^\dagger \tau^b N \right) \, , \label{TracelessSpurion:Identity} \\
        \left(N^\dagger \tilde Q_\pm \sigma^i N \right) \left( N^\dagger \tilde Q_\pm \sigma^i N \right) & = & c_{a, \pm} c_{b, \pm} \left( N^\dagger \tau^a \sigma^i N \right) \left( N^\dagger \tau^b \sigma^i N \right) \label{TracelessSpurion:Spin} \, .
    \end{eqnarray}
For $\Gamma = \tau^c$, 
    \begin{eqnarray}
        \left(N^\dagger \tilde Q_\pm \tau^c  N \right)  \left( N^\dagger \tilde Q_\pm \tau^c  N \right) & =&   c_{a, \pm} c_{a, \pm} \left( N^\dagger N \right)^2 - c_{a, \pm} c_{a, \pm} \left( N^\dagger \tau^b N \right)^2 \nonumber \\ && + c_{a, \pm} c_{b, \pm} \left( N^\dagger \tau^a N \right) \left( N^\dagger \tau^b N \right)\, ,
    \end{eqnarray}
and the operator $\left( N^\dagger \tau^b N \right)^2$ can be removed  using the Fierz identity of Eq.~\eqref{eq:NNFierz2} to obtain
    \begin{eqnarray}\label{tau.reduc}
        \left(N^\dagger \tilde Q_\pm \tau^c  N \right) \left( N^\dagger \tilde Q_\pm \tau^c  N \right)  &=&  3 c_{a, \pm} c_{a, \pm} \left( N^\dagger N \right)^2 + c_{a, \pm} c_{a, \pm} \left( N^\dagger \sigma^i N \right)^2 \nonumber \\ &&+ c_{a, \pm} c_{b, \pm} \left( N^\dagger \tau^a N \right) \left( N^\dagger \tau^b N \right)\, .
    \end{eqnarray}
The first two terms  in Eq.~\eqref{tau.reduc} have the same bilinear structure as the operators $\calO_{1,1}$ and $\calO_{1,2}$, respectively. Their contributions can be absorbed into a redefinition of the LECs of these operators.  The third term in Eq.~\eqref{tau.reduc} is  Eq.~\eqref{TracelessSpurion:Identity}. For $\Gamma = \sigma^i \tau^c$,  Eq.~\eqref{tau.reduc} appears again except that the last term  is Eq.~\eqref{TracelessSpurion:Spin} instead of Eq.~\eqref{TracelessSpurion:Identity}. 
This shows that $\Gamma = \tau^c$ and $\sigma^i\tau^c$ do not yield additional independent operators and can be neglected.

Additional relationships exist among some of the operators corresponding to $\Gamma=\1$ and $\Gamma=\sigma^i$. For $B_3$, $B_6$, and $B_{10}$, $\Gamma = \1$ can be eliminated by applying Fierz transformations to  Eq.~\eqref{TracelessSpurion:Identity} along with the decomposition in Eq.~\eqref{Traceless_isovector}. Using
    \begin{equation}
        \Tr( \tilde Q_\pm^2 ) = 2 c_{a, \pm} c_{a, \pm}\, ,
    \end{equation}
the Fierz transformation for Eq.~\eqref{TracelessSpurion:Identity} leads to
    \begin{multline}
        -3 \left[ \left( N^\dagger \tilde Q_\pm N \right) \left( N^\dagger \tilde Q_\pm N \right) - \frac{1}{6} \Tr( \tilde Q_\pm \tilde Q_\pm ) \left( N^\dagger \tau^a N \right)^2  \right] \\
        =  \left( N^\dagger \tilde Q_\pm N \right) \left( N^\dagger \sigma^i \tilde Q_\pm N \right) - \frac{1}{6} \Tr( \tilde Q_\pm \tilde Q_\pm ) \left( N^\dagger \sigma^i \tau^a N \right)^2\, , \label{eq:traceless_tensor_fierz}
    \end{multline}
which can be arranged, with the help of additional Fierz transformations, to be
    \begin{eqnarray}
        \left( N^\dagger \tilde Q_\pm N \right) \left( N^\dagger \tilde Q_\pm N \right) & = & - \frac{1}{3} \left( N^\dagger \tilde Q_\pm N \right) \left( N^\dagger \sigma^i \tilde Q_\pm N \right) - \frac{1}{2} \Tr( \tilde Q_\pm \tilde Q_\pm ) \left( N^\dagger N \right)^2 \nonumber \\
        & & - \frac{1}{6} \Tr( \tilde Q_\pm \tilde Q_\pm ) \left( N^\dagger \sigma^i N \right)^2 \, .
    \end{eqnarray}
Therefore,  the choice of $\Gamma = \1$ can be eliminated from $B_3$, $B_6$, and $B_{10}$ in favor of combinations of $\Gamma = \sigma^i$ and operators from $B_{11}$, $B_{12}$, and $B_{13}$.

Following the same procedure for operators from $B_2$, $B_5$, $B_8$, and $B_9$ 
results in 
    \begin{eqnarray}
        \Tr(Q_\pm) \left( N^\dagger \tilde Q_\pm N \right) \left( N^\dagger N \right) & = & c_{a, \pm} \Tr(Q_\pm) \left( N^\dagger N \right)^2 \, , \\
        \Tr(Q_\pm) \left( N^\dagger \tilde Q_\pm \sigma^i N \right) \left( N^\dagger \sigma^i N \right) & = & c_{a, \pm} \Tr(Q_\pm) \left( N^\dagger N \right)^2 \, ,
    \end{eqnarray}
for $\Gamma = \1$ and $\sigma^i$,  respectively. When $\Gamma = \tau^b$, Fierz transformations yield
    \begin{eqnarray}
        \Tr(Q_\pm) \left( N^\dagger \tilde Q_\pm \tau^b N \right) \left( N^\dagger \tau^b N \right) & = &  - c_{a, \pm} \Tr(Q_\pm) \left[ 2 \left( N^\dagger N \right)^2 + \left( N^\dagger \sigma^i N \right)^2 \right] \, .
    \end{eqnarray}
 Similarly, $\Gamma = \sigma^i \tau^b$ leads to
    \begin{eqnarray}
        \Tr(Q_\pm) \left( N^\dagger \tilde Q_\pm \sigma^i \tau^b N \right) \left( N^\dagger \sigma^i \tau^b N \right) & = &  -3 c_{a, \pm} \Tr(Q_\pm) \left( N^\dagger N \right)^2 \, . 
    \end{eqnarray}
This shows that again the operators with $\Gamma=\tau^a$ and $\Gamma=\sigma^i \tau^a$ are redundant for creating  a complete leading-in-\Nc description. As before, additional relationships exist between the $\Gamma=\1$ and $\Gamma = \sigma^i$ operators.
Eliminating the remaining redundancy through Fierz transformations leads to
    \begin{equation}
        -3 \left( N^\dagger N \right) \left( N^\dagger \tilde Q_\pm N \right) = \left( N^\dagger \sigma^i N \right) \left( N^\dagger \sigma^i \tilde Q_\pm N \right) \, .
    \end{equation}  
Therefore, either $\Gamma = \1$ or $\Gamma = \sigma^i$ may be retained, and both choices give the same large-\Nc counting.

This process eliminates the operators that possess a subleading spin-flavor structure. Any additional factors of \Nc that might be present will arise from pion fields in the expansion of $u$; however, these factors will not change the spin-flavor structure of the nucleon bilinears, and thus only lead to additional $1/\sqrt{\Nc}$ suppressions arising from factors of $1/F$.

\bibliography{references}
\end{document}